# METHODS IN *Mathematica* FOR SOLVING ORDINARY DIFFERENTIAL EQUATIONS

Ünal Göktaş[1] and Devendra Kapadia[2]
[1]Department of Computer Engineering, Turgut Özal University
Keçiören, Ankara 06010, Turkey. ugoktas@turgutozal.edu.tr
[2]Wolfram Research, Inc.
100 Trade Center Drive, Champaign, IL 61820, U.S.A. dkapadia@wolfram.com

**Abstract-** An overview of the solution methods for ordinary differential equations in the *Mathematica* function DSolve is presented.
**Key Words-** Differential Equation, *Mathematica*, Computer Algebra

## 1. INTRODUCTION

As they arise in the mathematical formulations of real-life problems, differential equations play a central role in displaying the interrelations between mathematics, physical and biological sciences or engineering [1]. Hence studying the solution methods of differential equationbs has always been an important research area [2, 3, 4]. With the invention of computer algebra systems, ongoing efforts for finding new methods for computation of solutions of differential equations led to exciting developments. An understanding of the scope and built-in algorithms in such systems is very useful while applying them in practice as they typically allow for a variety of approaches (symbolic, numerical, and graphical) for solving differential equations.

In this paper, we give an overview of available methods for solving ordinary differential equations (ODEs) in closed form and give examples of these methods in action as they are being used in DSolve, the function for solving differential equations in *Mathematica* [5], a major computer algebra system. In section 2, we give a list of methods for solving first-order ODEs. Section 3 contains the methods for solving second or higher order linear ODEs. In section 4, we deal with second or higher order ODEs which are nonlinear. Section 5 contains the methods for handling systems of ODEs. The examples have been chosen to illustrate the structure of typical members of each class of ODEs and we have tried to give insight into the key ideas and algorithms which are used to solve them. We have also included applications of differential equations to population dynamics and differential geometry. The final section offers suggestions for pursuing the study of differential equations in greater depth using *Mathematica*.

## 2. FIRST-ORDER ODEs

As they also become useful when solving higher order equations and systems of ODEs, studying the solution methods of first-order ODEs is really important. For some classes of first-order ODEs solution methods are known and well-studied [3, 4]. Among these classes of equations, we can list: linear, separable, Bernoulli, homogeneous, inverse linear, exact and Clairaut type first-order ODEs. Riccati, Abel [6] and Chini type first-order ODEs are also well-studied and many sub-classes of these equations for which a solution can be found are identified. For first-order ODEs which do not fit into one of these classes, one can try: recognizing the defining differential equations of the special functions of physics, finding integrating factors or computing Lie symmetries of the ODE [7, 8, 9].

In the well-known books by Kamke [3] and Murphy [4], the standard methods are covered. When dealing with a first-order ODE, based on the statistics of equations listed in [3], one can choose the sequence of the methods to apply. A great portion of the 576 first-order ODEs listed in [3] can be solved with the standard methods for solving linear, separable, Riccati and Abel type equations. For example, 103 equations fit into the separable class.

## 2.1. Linear equations

Linear first-order ODEs are identified as $y'(x) = a(x)y(x) + b(x)$ and the solution is given after integration:

```
In[2]:= DSolve[y'[x] == a[x] y[x] + b[x], y[x], x]
```

$$\text{Out[2]}= \left\{\left\{y[x] \to e^{\int_1^x a[K[1]]\, dK[1]} C[1] + e^{\int_1^x a[K[1]]\, dK[1]} \int_1^x e^{-\int_1^{K[2]} a[K[1]]\, dK[1]} b[K[2]]\, dK[2]\right\}\right\}$$

In the above solution, $K[1]$ and $K[2]$ denote the dummy integration variables. To suppress possible messages generated by DSolve, we initially ran the following command:

```
In[1]:= Off[InverseFunction::ifun, Solve::ifun, Solve::svars, Solve::tdep]
```

## 2.2. Separable equations

First-order ODEs which can be written as $y'(x) = f(x)g(y(x))$ are called separable equations and the solution is again given after integration:

```
In[3]:= DSolve[y'[x] == f[x] g[y[x]], y[x], x]
```

$$\text{Out[3]}= \left\{\left\{y[x] \to \text{InverseFunction}\left[\int_1^{\#1} \frac{1}{g[K[1]]}\, dK[1]\, \&\right]\left[C[1] + \int_1^x f[K[2]]\, dK[2]\right]\right\}\right\}$$

The logistic equation $y'(x) = y(x)(1 - b\, y(x))$ is a well-known separable equation. Now let us solve the logistic equation with the initial condition $y(0) = a$:

```
In[4]:= solutionofLogisticequation = DSolve[{y'[x] == y[x] (1 - b y[x]), y[0] == a}, y[x], x]
```

$$\text{Out[4]}= \left\{\left\{y[x] \to \frac{a\, e^x}{1 - a b + a b\, e^x}\right\}\right\}$$

Now let us plot the solution we found with various settings of the variables $a$ and $b$:

```
In[5]:= Plot[Evaluate[y[x] /. solutionofLogisticequation /. {b -> 1/27} /.
         {{a -> 1/13}, {a -> 1/2}, {a -> 1/6}, {a -> 1/28}}], {x, 0, 18}, PlotStyle -> Thickness[0.01]]
```

Methods in Mathematica for Solving Ordinary Differential Equations

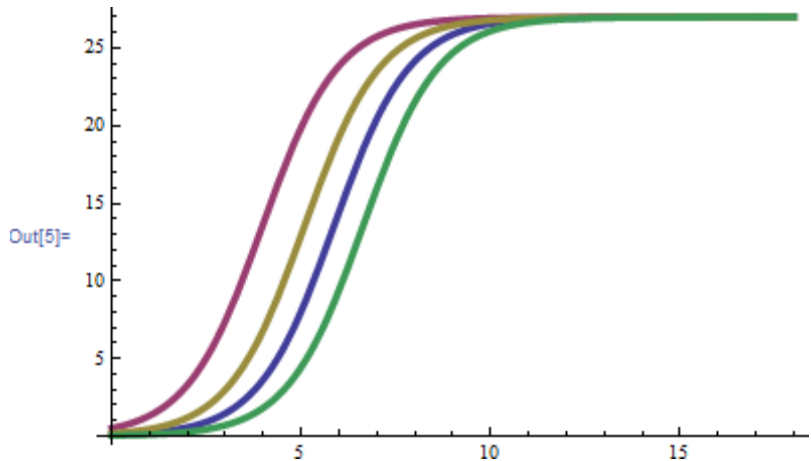

Out[5]=

### 2.3. Bernoulli type equations

Equations of the form $y'(x) = f(x)y(x) + g(x)y(x)^k$ are called the Bernoulli type equations and the solution is found after integration. Following example is the equation 1.34 from [3]:

In[6]:= $\mathbf{DSolve}\left[g[x]\,y[x] + f[x]\,y[x]^2 + y'[x] = 0,\ y[x],\ x\right]$

Out[6]= $\left\{\left\{y[x] \to \dfrac{e^{\int_1^x -g[K[1]]\,dK[1]}}{C[1] - \int_1^x e^{\int_1^{K[2]} -g[K[1]]\,dK[1]}\,f[K[2]]\,dK[2]}\right\}\right\}$

### 2.4. Homogeneous equations

A first-order ODE of the form $y'(x) = f(x, y(x))$ is called homogeneous if the substitution $u(x) = y(x)/x$ reduces it to a separable equation [4]:

In[7]:= $\mathbf{DSolve}\left[y'[x] = \dfrac{-4xy[x] - 2\sqrt{2x^2\,y[x]^2 + y[x]^4}}{2\,y[x]^2},\ y[x],\ x\right]$

Out[7]= $\left\{\left\{y[x] \to -\sqrt{e^{2C[1]} - 2e^{C[1]}x - x^2}\right\},\ \left\{y[x] \to \sqrt{e^{2C[1]} - 2e^{C[1]}x - x^2}\right\}\right\}$

### 2.5. Inverse linear equations

ODEs of the form $y'(x) = f(x, y(x))$ are called inverse linear if $\dfrac{1}{f(x,y(x))}$ becomes linear after substituting $x = x(y)$ and $y(x) = y$:

In[8]:= $\mathbf{DSolve}\left[y'[x] = \dfrac{1}{x - y[x]},\ y,\ x\right]$

Out[8]= $\left\{\left\{y \to \text{Function}\left[\{x\},\ -1 + x - \text{ProductLog}\left[e^{-1+x}\,C[1]\right]\right]\right\}\right\}$

### 2.6. Riccati type equations

First-order ODEs of the form $y'(x) = a(x) + b(x)y(x) + c(x)y(x)^2$ are called Riccati type equations. The methods for solving the Riccati type equations are studied in [3, 4]:

In[9]:= `DSolve[y'[x] == x^3 + y[x]^2, y[x], x]`

Out[9]= $\left\{\left\{y[x] \to \left(-\text{BesselJ}\left[-\frac{1}{5}, \frac{2x^{5/2}}{5}\right]C[1] + x^{5/2}\left(-2\,\text{BesselJ}\left[-\frac{4}{5}, \frac{2x^{5/2}}{5}\right] - \text{BesselJ}\left[-\frac{6}{5}, \frac{2x^{5/2}}{5}\right]C[1] + \text{BesselJ}\left[\frac{4}{5}, \frac{2x^{5/2}}{5}\right]C[1]\right)\right) \Big/ \left(2x\left(\text{BesselJ}\left[\frac{1}{5}, \frac{2x^{5/2}}{5}\right] + \text{BesselJ}\left[-\frac{1}{5}, \frac{2x^{5/2}}{5}\right]C[1]\right)\right)\right\}\right\}$

The following Riccati type ODE is first transformed into a linear second-order ODE and once the second-order ODE is solved, solution to the original first-order Riccati type ODE is found by inverse transformation:

In[10]:= `DSolve[y'[x] == 2 - x y[x]^2, y[x], x]`

Out[10]= $\left\{\left\{y[x] \to \frac{2^{2/3}\,x\,\text{AiryBi}[2^{1/3}\,x] + 2^{2/3}\,x\,\text{AiryAi}[2^{1/3}\,x]\,C[1]}{x\,(\text{AiryBiPrime}[2^{1/3}\,x] + \text{AiryAiPrime}[2^{1/3}\,x]\,C[1])}\right\}\right\}$

Note that, in the above general solution if an initial condition such as $y(0) = 1$ has to be satisfied, this could be handled by using limit:

In[11]:= `Solve[Limit[y[x] /. %[[1]], x -> 0] == 1, C[1]]`

Out[11]= $\left\{\left\{C[1] \to \frac{-6^{2/3}\,\text{Gamma}\left[\frac{1}{3}\right] + 3\,\text{Gamma}\left[\frac{2}{3}\right]}{2^{2/3}\,3^{1/6}\,\text{Gamma}\left[\frac{1}{3}\right] + \sqrt{3}\,\text{Gamma}\left[\frac{2}{3}\right]}\right\}\right\}$

### 2.7. Chini type equations

Equations of the form $y'(x) = f(x)y(x)^n + g(x)y(x) + h(x)$ are called Chini type first-order ODEs and various cases when a solution is directly formed are given in [3, 4]:

In[12]:= `DSolve[y'[x] == 5 y[x]^4 + 3 x^{-4/3}, y[x], x]`

Out[12]= $\text{Solve}\left[-45\,\text{RootSum}\left[-45 + 3^{1/4}\,5^{3/4}\,\#1 - 45\,\#1^4\,\&,\,\frac{\text{Log}\left[-\#1 + \left(\frac{5}{3}\right)^{1/4}(x^{4/3})^{1/4}\,y[x]\right]}{3^{1/4}\,5^{3/4} - 180\,\#1^3}\,\&\right] == C[1] + \frac{3^{3/4}\,5^{1/4}\,(x^{4/3})^{1/4}\,\text{Log}[x]}{x^{1/3}},\,y[x]\right]$

### 2.8. Abel type equations

Equations of the form $y'(x) = a(x) + b(x)y(x) + c(x)y(x)^2 + d(x)y(x)^3$ are called Abel equations of the first kind. Abel equations of the second kind are of the form $y'(x) = \frac{a(x) + b(x)y(x) + c(x)y(x)^2 + d(x)y(x)^3}{e(x) + f(x)y(x)}$. The cases where direct solution methods are available are studied in [3, 4] and [6]:

In[13]:= `DSolve[y'[x] == -a y[x]^2 - x y[x]^3, y, x]`

Out[13]= $\text{Solve}\left[-\frac{1}{2}a^2\left(\frac{2\,\text{ArcTan}\left[\frac{-a-2xy[x]}{\sqrt{-1-\frac{4}{a^2}}\,a}\right]}{\sqrt{-1-\frac{4}{a^2}}} - \text{Log}\left[\frac{-1-xy[x](-a-xy[x])}{x^2 y[x]^2}\right]\right) == C[1] - a^2 \text{Log}[x],\ y[x]\right]$

To verify this implicit solution, we can do:

In[14]:= `Together[∂ₓ %[[1]] /. y'[x] → -a y[x]^2 - x y[x]^3]`

Out[14]= True

### 2.9. Exact equations (and computing integrating factors)

A first-order ODE is exact if $y'(x) - f(x, y(x)) = \tfrac{d}{dx} R(x, y(x))$. For an ODE $y'(x) = f(x, y(x))$, if $\mu(x, y(x))(y'(x) - f(x, y(x))) = \tfrac{d}{dx} R(x, y(x))$, then $\mu(x, y(x))$ is an integrating factor. This means that you make the ODE exact if you can find an integrating factor [3, 4]:

In[15]:= `DSolve[x + f[x^2 + y[x]^2] g[x] + y[x] y'[x] == 0, y, x]`

Out[15]= $\text{Solve}\left[\int_1^x \left(g[K[1]] + \frac{K[1]}{f[K[1]^2 + y[x]^2]}\right) dK[1] + \int_1^{y[x]} \frac{K[2]}{f[1 + K[2]^2]} dK[2] == C[1],\ y[x]\right]$

### 2.10. Lie symmetry methods

The knowledge of a symmetry in the form of an infinitesimal generator reduces the order of an ODE which typically simplifies the problem. DSolve function checks for standard types of symmetries in the given ODE and uses them to return the solution [7, 8, 9].

For the following ODE 1.357 of [3], DSolve uses the symmetry methods to find the solution:

In[16]:= `DSolve[Cos[y[x]] (1 - x Cos[y[x]]) + x Log[x] Sin[y[x]] y'[x] == 0, y[x], x]`

Out[16]= $\left\{\left\{y[x] \to -\text{ArcSec}\left[\frac{x - C[1]}{\text{Log}[x]}\right]\right\}, \left\{y[x] \to \text{ArcSec}\left[\frac{x - C[1]}{\text{Log}[x]}\right]\right\}\right\}$

For the ODE 1.188 of [3], DSolve returns an implicit solution:

In[17]:= `DSolve[-b x^(3n) - a y[x]^3 + x^(1+2n) y'[x] == 0, y[x], x]`

Out[17]= $\text{Solve}\left[-\text{RootSum}\left[-1 + \left(\frac{n^3}{a b^2}\right)^{1/3} \#1 - \#1^3\ \&,\ \frac{\text{Log}\left[-\#1 + \left(\frac{a x^{-3n}}{b}\right)^{1/3} y[x]\right]}{\left(\frac{n^3}{a b^2}\right)^{1/3} - 3\#1^2}\ \&\right] ==\right.$

$\left. C[1] + b x^n \left(\frac{a x^{-3n}}{b}\right)^{1/3} \text{Log}[x],\ y[x]\right]$

### 2.11. Clairaut equations

ODEs of the form $y(x) = x\,y'(x) + f(y'(x))$ are called Clairaut equations [3]:

```
In[18]:= DSolve[y[x] == x y'[x] + f[y'[x]], y, x]
Out[18]= {{y → Function[{x}, x C[1] + f[C[1]]]}}
```

**2.12. Recognizing defining equations of special functions**

Several elliptic functions are defined by first-order ODEs as illustrated below for the WeierstrassPPrime function:

```
In[19]:= DSolve[2 y'[x]^3 - 3 y'[x]^2 - 54 y[x]^2 - 27 y[x]^4 - 26 == 0, y[x], x]
Out[19]= {{y[x] → WeierstrassPPrime[x + C[1], {1, 1}]}}
```

## 3. LINEAR SECOND OR HIGHER ORDER ODEs

There are various standard methods for solving linear second or higher order ODEs. Equations with constant coefficients are solved using the roots of the characteristic equation. The Euler-Legendre type equations can be transformed into equations with constant coefficients. Some equations are exact and so can be directly integrated whereas some can be directly integrated after multiplying with an integrating factor. For second-order equations with rational coefficients there is the well-known Kovacic algorithm for finding the Liouvillian solutions [10]. For second or higher order equations with rational coefficients, in [11, 12] and [13] the algorithms for finding rational and exponential solutions are given. The hypergeometric PFQ type solutions are found using the algorithms in [14] and [15]. In [16] an algorithm using the concept of symmetric powers, and in [17] an algorithm using the concept of symmetric products are given. There are also various factorization algorithms for higher order ODEs in [18] and [19].

**3.1. ODEs with constant coefficients**

The general solution of linear ODEs with constant coefficients are found by using the roots of the characteristic equation for the ODE:

```
In[20]:= DSolve[y''[x] + 2 y'[x] + y[x] == 0, y[x], x]
Out[20]= {{y[x] → e^-x C[1] + e^-x x C[2]}}
```

**3.2. Euler-Legendre type equations**

The following equation is an Euler-Legendre ODE which can be solved by transforming it to a linear ODE with constant coefficients:

```
In[21]:= DSolve[x^2 y''[x] - 13 x y'[x] + y[x] == 0, y[x], x]
Out[21]= {{y[x] → x^(7-4√3) C[1] + x^(7+4√3) C[2]}}
```

**3.3. Exact equations (and integrating factors)**

The following is an example of an exact ODE since the left-hand side can be integrated to a first-order expression whose solution gives one element of the general solution of the second-order ODE:

```
In[22]:= secondorder = 2 x y[x] + x^2 y'[x] + y''[x];
```

```
In[23]:= firstorder = Integrate[secondorder, x]
```
Out[23]= $x^2 y[x] + y'[x]$

```
In[24]:= DSolve[firstorder == 0, y[x], x]
```
Out[24]= $\{\{y[x] \to e^{-\frac{x^3}{3}} C[1]\}\}$

```
In[25]:= DSolve[secondorder == 0, y[x], x]
```
Out[25]= $\{\{y[x] \to e^{-\frac{x^3}{3}} C[2] - \frac{e^{-\frac{x^3}{3}} x C[1] \text{Gamma}[\frac{1}{3}, -\frac{x^3}{3}]}{3^{2/3} (-x^3)^{1/3}}\}\}$

### 3.4. Recognizing the defining equations of special functions

There are many physical processes which are modeled by linear second or higher order ODEs. DSolve can recognize the defining ODEs for many special functions:

```
In[26]:= DSolve[w''[x] - x w[x] == 0, w[x], x]
```
Out[26]= $\{\{w[x] \to \text{AiryAi}[x] C[1] + \text{AiryBi}[x] C[2]\}\}$

```
In[27]:= DSolve[(a x + b)^2 y''[x] + 2 a (a x + b) y'[x] + a^2 ((a x + b)^2 - n (n + 1)) y[x] == 0, y[x], x]
```
Out[27]= $\{\{y[x] \to C[1] \text{SphericalBesselJ}[n, b + a x] + C[2] \text{SphericalBesselY}[n, b + a x]\}\}$

```
In[28]:= DSolve[y''[x] + (-\frac{1}{4} + \frac{k}{x} + \frac{\frac{1}{4} - m^2}{x^2}) y[x] == 0, y[x], x]
```
Out[28]= $\{\{y[x] \to C[1] \text{WhittakerM}[k, m, x] + C[2] \text{WhittakerW}[k, m, x]\}\}$

```
In[29]:= DSolve[\frac{(b x + c) y''[x]}{b^2} + \frac{(a + 1 - (b x + c)) y'[x]}{b} + n y[x] == 0, y[x], x]
```
Out[29]= $\{\{y[x] \to C[1] \text{HypergeometricU}[-n, 1 + a, c + b x] + C[2] \text{LaguerreL}[n, a, c + b x]\}\}$

```
In[30]:= DSolve[(g^2 - g^2 x^2 - \frac{m^2}{1-x^2} + \text{SpheroidalEigenvalue}[n, m, g]) y[x] - 2 x y'[x] + (1 - x^2) y''[x] == 0, y[x], x]
```
Out[30]= $\{\{y[x] \to C[1] \text{SpheroidalPS}[n, m, g, x] + C[2] \text{SpheroidalQS}[n, m, g, x]\}\}$

```
In[31]:= DSolve[y''[x] - (a - 2 q \text{Cosh}[2 x]) y[x] == 0, y[x], x]
```
Out[31]= $\{\{y[x] \to C[1] \text{MathieuC}[a, q, i x] - C[2] \text{MathieuS}[a, q, i x]\}\}$

```
In[32]:= DSolve[x^4 y^{(4)}[x] + 2 x^3 y^{(3)}[x] - (1 + 2 v^2)(x^2 y''[x] - x y'[x]) + (v^4 - 4 v^2 + x^4) y[x] == 0, y[x], x]
```
Out[32]= $\{\{y[x] \to C[4] \text{KelvinBei}[v, x] + C[3] \text{KelvinBer}[v, x] + C[2] \text{KelvinKei}[v, x] + C[1] \text{KelvinKer}[v, x]\}\}$

### 3.5. Kovacic algorithm

This is a standard algorithm for solving second-order linear homogeneous ODEs with rational function coefficients [10]:

In[33]:= `DSolve[x y''[x] + (10 x^3 - 1) y'[x] + 5 x^2 (5 x^3 + 1) y[x] == 0, y[x], x]`

Out[33]= $\left\{\left\{y[x] \to e^{-\frac{1}{3}(5x^3)} C[1] + \frac{1}{2} e^{-\frac{1}{3}(5x^3)} x^2 C[2]\right\}\right\}$

In the above solution, the piece with C[1] is found by the Kovacic algorithm, and the second independent solution is found by reduction of order.

### 3.6. Rational and exponential solutions

For finding rational solutions of linear ODEs with rational coefficients the algorithms in [11] and [12], and for finding exponential solutions the algorithm in [13] can be used. Here is a nice example where we can see a nice harmony of the methods for rational solutions, reduction of order and recognizing special functions:

In[34]:= `DSolve[`
$$\frac{(6 - 24x + 4x^2 - 10x^3 + 9x^4 - 4x^5 + x^6) y[x]}{(-1+x)^4 x^2} + \frac{(-8 + 30x - 10x^2 + 10x^3 - 3x^4 - 2x^5 + x^6) y'[x]}{(-1+x)^3 x^2} +$$
$$\frac{(4 - 16x + 5x^2 - 2x^3) y''[x]}{(-1+x)^2 x^2} + \frac{(4 - x^2) y^{(3)}[x]}{(-1+x) x} == 0, y[x], x\bigg]$$

Out[34]= $\left\{\left\{y[x] \to \frac{(-1+x) C[1]}{-4+x^2} + \frac{(-1+x) \text{AiryAi}[x] C[2]}{-4+x^2} + \frac{(-1+x) \text{AiryBi}[x] C[3]}{-4+x^2}\right\}\right\}$

### 3.6. Hypergeometric PFQ type solutions

Hypergeometric PFQ type solutions are found by using the algorithms in [14] and [15]. The following example is equation 2.16 from [3] with $a = b = 1, c = 2$:

In[35]:= `DSolve[y''[x] + (x^4 + x) y[x] == 0, y[x], x]`

Out[35]= $\left\{\left\{y[x] \to \frac{2^{1/3} e^{\frac{ix^3}{3}} (x^3)^{1/3} C[1] \text{HypergeometricU}\left[\frac{1}{3} - \frac{i}{6}, \frac{2}{3}, -\frac{2ix^3}{3}\right]}{x} + \right.\right.$
$$\left.\left.\frac{2^{1/3} e^{\frac{ix^3}{3}} (x^3)^{1/3} C[2] \text{LaguerreL}\left[-\frac{1}{3} + \frac{i}{6}, -\frac{1}{3}, -\frac{2ix^3}{3}\right]}{x}\right\}\right\}$

### 3.7. Symmetric power and symmetric product solutions

Given a second-order linear ODE with basis $\{u, v\}$ for its general solution, we can construct an ODE of order $n (> 2)$ whose general solution has basis $\{u^{n-1}, u^{n-2}v, \ldots, v^{n-1}\}$. This higher order equation is called the $(n-1)$th symmetric power of the second-order ODE [16]. Following example is a third order ODE which is solved using the second symmetric power of the Legendre's equation:

In[36]:= `DSolve[` $-\frac{4n(1+n) x y[x]}{(1-x^2)^2} + \left(\frac{4x^2}{(1-x^2)^2} - \frac{2}{1-x^2} + \frac{4n(1+n)}{1-x^2}\right) y'[x] - \frac{6 x y''[x]}{1-x^2} + y^{(3)}[x] == 0, y[x], x$ `]`

Out[36]= $\{\{y[x] \to C[1] \text{LegendreP}[n, x]^2 + C[2] \text{LegendreP}[n, x] \text{LegendreQ}[n, x] + C[3] \text{LegendreQ}[n, x]^2\}\}$

Given a pair of second-order linear ODEs with the bases $\{r, s\}$ and $\{u, v\}$, it is possible to construct a fourth order ODE whose general solution has the basis $\{ru, rv, su, sv\}$. This fourth order ODE is called the symmetric product of the second-order equations [17]:

```
In[37]:= DSolve[16 x^2 y''[x] + 8 x y'[x] + (1 - 16 x) y[x] == 0, y[x], x]
```

$$Out[37]= \{\{y[x] \to (-1)^{1/4} x^{1/4} \text{BesselI}[0, 2\sqrt{x}] C[1] + 2 x^{1/4} \text{BesselK}[0, 2\sqrt{x}] C[2]\}\}$$

```
In[38]:= DSolve[y''[x] - y[x] == 0, y[x], x]
```

$$Out[38]= \{\{y[x] \to e^x C[1] + e^{-x} C[2]\}\}$$

Here is the solution of the symmetric product of these ODEs:

$$In[39]:= \text{DSolve}\Big[y^{(4)}[x] + \frac{4(x-1)y^{(3)}[x]}{x(-1+2x)} - \frac{2(2x^3+x^2+1)y''[x]}{x^2(-1+2x)} - \frac{4(x^2-1-2x)y'[x]}{x^2(-1+2x)} +$$

$$\frac{(2x^3-5x^2+6x-1)y[x]}{x^2(-1+2x)} == 0, y[x], x\Big]$$

$$Out[39]= \{\{y[x] \to (-1)^{1/4} e^{-x} \text{BesselI}[0, 2\sqrt{x}] C[1] + \frac{1}{2}(-1)^{1/4} e^x \text{BesselI}[0, 2\sqrt{x}] C[2] +$$

$$2 e^{-x} \text{BesselK}[0, 2\sqrt{x}] C[3] + e^x \text{BesselK}[0, 2\sqrt{x}] C[4]\}\}$$

**3.8. Factorization**

DSolve has the implementations of factorization algorithms in [18] and [19]:

```
In[40]:= DSolve[y^(4)[x] - 2 x y''[x] - 2 y'[x] + x^2 y[x] == 0, y[x], x]
```

$$Out[40]= \{\{y[x] \to \text{AiryAi}[x] C[1] + \text{AiryBi}[x] C[2] +$$

$$(-\pi \text{AiryAiPrime}[x]^2 \text{AiryBi}[x] + \pi \text{AiryAi}[x] \text{AiryAiPrime}[x] \text{AiryBiPrime}[x]) C[3] +$$

$$(-\pi \text{AiryAiPrime}[x] \text{AiryBi}[x] \text{AiryBiPrime}[x] + \pi \text{AiryAi}[x] \text{AiryBiPrime}[x]^2) C[4]\}\}$$

**3.9. Equations solved after a transformation**

DSolve uses a number of transformation rules in order to solve ODEs whose coefficients are rational in either trigonometric, hyperbolic or exponential functions. The basic idea is to transform the given ODE into one whose coefficients are rational in the new independent variable.

In the following example, we use DSolve to find quantum eigenfunctions for a modified Pöschl–Teller potential, which requires the solution of a linear second-order ODE with hyperbolic coefficients:

```
In[41]:= DSolve[y''[x] + j (j+1) Sech[x]^2 y[x] == n^2 y[x], y[x], x]
```

$$Out[41]= \{\{y[x] \to C[1] \text{LegendreP}[j, n, \text{Tanh}[x]] + C[2] \text{LegendreQ}[j, n, \text{Tanh}[x]]\}\}$$

## 4. NONLINEAR SECOND OR HIGHER ORDER ODEs

DSolve has special methods for solving important classes of nonlinear ODEs which arise in applications. This includes the standard methods for handling the equations with missing variables, exact equations, and homogeneous equations [3, 4], computing integrating factors due to the algorithms in [20], and computing the Lie symmetries [9] and applying transformation rules that reduce the equation to one of the standard types.

### 4.1. Equations with missing variables

In the following example, the differential equation does not depend explicitly on the independent variable $x$, and a first integral can be found after making the substitution $p = y'(x)$:

```
In[42]:= DSolve[y''[x] == Cos[y[x]], y[x], x]
```

$$Out[42]= \left\{\left\{y[x] \to \frac{1}{2}\left(\pi - 4\,\text{JacobiAmplitude}\left[\frac{1}{2}\sqrt{(2+C[1])(x+C[2])^2}\,,\,\frac{4}{2+C[1]}\right]\right)\right\},\right.$$
$$\left.\left\{y[x] \to \frac{1}{2}\left(\pi + 4\,\text{JacobiAmplitude}\left[\frac{1}{2}\sqrt{(2+C[1])(x+C[2])^2}\,,\,\frac{4}{2+C[1]}\right]\right)\right\}\right\}$$

### 4.2. Transformations

It is sometimes possible to remove the explicit dependence of an ODE on the independent variable by using a transformation. For instance, the following ODE is solved by making the transformation $y(x) = u(x) + x$, which leads to a differential equation for $u(x)$ with missing independent variable $x$:

```
In[43]:= DSolve[y''[x] == Exp[y[x] - x], y[x], x]
```

$$Out[43]= \left\{\left\{y[x] \to x + \text{Log}\left[\frac{1}{2}C[1]\left(-1 + \text{Tanh}\left[\frac{1}{2}\sqrt{C[1](x+C[2])^2}\,\right]^2\right)\right]\right\}\right\}$$

### 4.3. Homogeneous equations

The left-hand side of the following nonlinear ODE is a homogeneous function of degree 2 in the variables $\{y(x), y'(x), y''(x)\}$, and the substitution $y(x) = \exp(\int u(x)\,dx)$ reduces the problem to solving a first-order differential equation in $u(x)$:

```
In[44]:= DSolve[7 y[x] y''[x] - 11 y'[x]^2 == 0, y[x], x]
```

$$Out[44]= \left\{\left\{y[x] \to \frac{C[2]}{(4x + 7C[1])^{7/4}}\right\}\right\}$$

### 4.3. Exact equations (and integrating factors)

In [20] methods for finding integrating factors of the form $\mu(x, y(x))$, $\mu(x, y'(x))$, and $\mu(y(x), y'(x))$ are presented. $\mu = \frac{1}{x\,y'(x)^2}$ is used as an integrating factor for solving the following ODE:

```
In[45]:= DSolve[y''[x] + y'[x]/x - Sin[y[x]] y'[x]^2 x - Cos[y[x]] x^2 y'[x]^3 == 0, y[x], x]
```

$$Out[45]= \text{Solve}\left[-\frac{e^{C[1]y[x]}}{xC[1]} + \frac{e^{C[1]y[x]}(-\text{Cos}[y[x]] + C[1]\,\text{Sin}[y[x]])}{C[1](1+C[1]^2)} == C[2],\ y[x]\right]$$

## 5. SYSTEMS OF ODEs

DSolve has a variety of methods for solving systems of ODEs with constant or variable coefficients. Systems with higher-order derivatives are internally reduced to first-order systems and, wherever possible, the system is decoupled to reduce the problem to solving a set of independent single ODEs. We will now give a few examples for solving systems of ODEs with increasing levels of complexity to illustrate the techniques used for solving them.

### 5.1. Systems with constant coefficients

Linear systems with constant coefficients are solved by applying matrix exponentiation to the coefficient matrix:

In[46]:= `DSolve[{x'[t] == -2 x[t] + 3 y[t], y'[t] == 2 x[t] - y[t]}, {x[t], y[t]}, t]`

Out[46]= $\left\{\left\{x[t] \to \frac{1}{5} e^{-4t}(3 + 2 e^{5t}) C[1] + \frac{3}{5} e^{-4t}(-1 + e^{5t}) C[2],\right.\right.$
$\left.\left. y[t] \to \frac{2}{5} e^{-4t}(-1 + e^{5t}) C[1] + \frac{1}{5} e^{-4t}(2 + 3 e^{5t}) C[2]\right\}\right\}$

### 5.2. Systems of decoupled equations

The following example illustrates a system composed of decoupled equations in which each equation involves a single dependent variable only. In such cases, the equations are solved independently using the available methods for single ODEs.

In[47]:= `DSolve[{y'[x] == x y[x], u''[x] == x u'[x]}, {y[x], u[x]}, x]`

Out[47]= $\left\{\left\{y[x] \to e^{\frac{x^2}{2}} C[1], u[x] \to C[3] + \sqrt{\frac{\pi}{2}} C[2] \operatorname{Erfi}\left[\frac{x}{\sqrt{2}}\right]\right\}\right\}$

### 5.3. Systems solved by one at a time integration

The following system of ODEs is solved by integrating the first equation which involves only $x(t)$ and then substituting the solution for $x(t)$ in the second equation to obtain a closed-form solution for $y(t)$:

In[48]:= $\text{DSolve}\left[\left\{x'[t] == x[t] \cos[t], y'[t] == \frac{x[t]}{e^{\sin[t]}}\right\}, \{x[t], y[t]\}, t\right]$

Out[48]= $\{\{x[t] \to e^{\sin[t]} C[1], y[t] \to t C[1] + C[2]\}\}$

### 5.4. Systems with patterns

The following linear system of ODEs is solved in closed form as the coefficient matrix has a special structure:

In[49]:= `DSolve[{x'[t] == f[t] x[t] + g[t] y[t], y'[t] == -g[t] x[t] + f[t] y[t]}, {x[t], y[t]}, t]`

Out[49]= $\left\{\left\{x[t] \to e^{\int_1^t f[K[2]] dK[2]} C[1] \cos\left[\int_1^t g[K[1]] dK[1]\right] + e^{\int_1^t f[K[2]] dK[2]} C[2] \sin\left[\int_1^t g[K[1]] dK[1]\right],\right.\right.$
$\left.\left. y[t] \to e^{\int_1^t f[K[2]] dK[2]} C[2] \cos\left[\int_1^t g[K[1]] dK[1]\right] - e^{\int_1^t f[K[2]] dK[2]} C[1] \sin\left[\int_1^t g[K[1]] dK[1]\right]\right\}\right\}$

### 5.5. Rational solutions

For linear systems with rational coefficients, the algorithms in [11, 12] for finding rational solutions are used:

$$\text{In[50]:= DSolve}\left[\left\{-\frac{(5+x)v[x]}{(-3-2x+x^2)(-1+x^3)} - \frac{(6+2x-3x^3-x^5)y[x]}{x(-3-2x+x^2)(-1+x^3)} + y'[x] = 0,\right.\right.$$

$$\left.\left. -\frac{(1+20x^2+3x^3)v[x]}{(5+x)(-1+x^3)} + \frac{4x(-3-2x+x^2)y[x]}{(5+x)(-1+x^3)} + v'[x] = 0\right\}, \{y[x], v[x]\}, x\right]$$

$$\text{Out[50]= }\left\{\left\{v[x] \to -\frac{C[1]}{3(5+x)} - \frac{x^4 C[2]}{3(5+x)}, y[x] \to -\frac{x C[1]}{3(-3+x)(1+x)} - \frac{x^2 C[2]}{3(-3+x)(1+x)}\right\}\right\}$$

### 5.6. Application: Solving the equations for a Cornu spiral

The following equations describe a Cornu spiral. This is a plane curve through the origin whose curvature is equal to the parameter value $s$ at every point.

In[51]:= **PlaneCurve** = {x'[s] = Cos[t[s]], y'[s] = Sin[t[s]], t'[s] = s, x[0] = 0, y[0] = 0, t[0] = 0};

The solution of the ODEs contains Fresnel functions.

In[52]:= **sol = DSolve[PlaneCurve, {x[s], y[s], t[s]}, s]**

Out[52]= $\left\{\left\{t[s] \to \frac{s^2}{2}, x[s] \to \sqrt{\pi}\ \text{FresnelC}\left[\frac{s}{\sqrt{\pi}}\right], y[s] \to \sqrt{\pi}\ \text{FresnelS}\left[\frac{s}{\sqrt{\pi}}\right]\right\}\right\}$

A graph of the curve clearly shows that the curvature becomes large as $s$ approaches infinity:

In[53]:= **ParametricPlot[Evaluate[{x[s], y[s]} /.sol[[1]]], {s, -20, 20}, PlotStyle → Red,
PlotPoints → 30, Background → Yellow]**

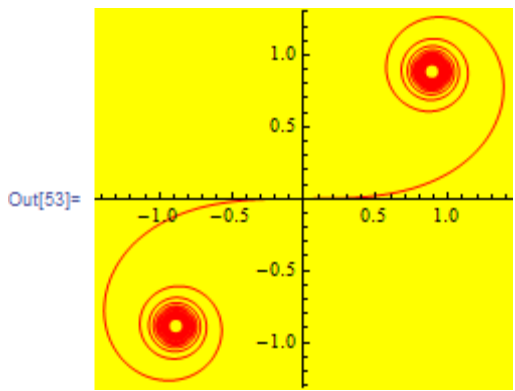

Out[53]=

## 6. CONCLUSION

Further information on DSolve is available on the documentation page for this function [21]. The tutorial [22] gives a comprehensive overview of the functionality available in DSolve for symbolic solutions of differential equations, along with further references for this topic. Numerical solutions of differential equations can be computed using the NDSolve function [23] in *Mathematica*.

We hope that the examples and ideas outlined in this paper will be useful for elementary and advanced courses on ordinary differential equations, as well as for solving differential equations which occur in research and design problems in practice.

Methods in Mathematica for Solving Ordinary Differential Equations